\documentclass[twocolumn]{aastex62}
\usepackage{amsmath}
\usepackage{float}
\usepackage{enumitem}
\usepackage{CJKutf8}
\newcommand{\chinesename}{{\begin{CJK}{UTF8}{gbsn}(王加冕)\end{CJK}}}

\begin{document}

\title{\uppercase{The Measured Impact of Chromatic Atmospheric Effects on Barycentric Corrections: Results from the EXtreme PREcision Spectrograph}}

\correspondingauthor{Ryan T. Blackman} 
\email{ryan.blackman@yale.edu}
\author[0000-0002-0303-3276]{Ryan T. Blackman} 
\author[0000-0001-7664-648X]{J. M. Joel Ong \chinesename}
\author[0000-0003-2221-0861]{Debra A. Fischer}
\affiliation{Department of Astronomy, Yale University, 52 Hillhouse Ave, New Haven, CT 06511, USA}

\keywords{atmospheric effects - techniques: radial velocities - instrumentation: spectrographs}

\begin{abstract}
One source of error in high-precision radial velocity measurements of exoplanet host stars is chromatic change in Earth's atmospheric transmission during observations. Mitigation of this error requires that the photon-weighted barycentric correction be applied as a function of wavelength across the stellar spectrum. We have designed a system for chromatic photon-weighted barycentric corrections with the EXtreme PREcision Spectrograph (EXPRES) and present results from the first year of operations, based on radial velocity measurements of more than $10^3$ high-resolution stellar spectra. For observation times longer than 250 seconds, we find that if the chromatic component of the barycentric corrections is ignored, a range of radial velocity errors up to 1 m s$^{-1}$ can be incurred with cross-correlation, depending on the nightly atmospheric conditions. For this distribution of errors, the standard deviation is 8.4 cm s$^{-1}$ for G-type stars, 8.5 cm s$^{-1}$ for K-type stars, and 2.1 cm s$^{-1}$ for M-type stars. This error is reduced to well-below the instrumental and photon-noise limited floor by frequent flux sampling of the observed star with a low-resolution exposure meter spectrograph.
\end{abstract}

\section{Introduction}
Exoplanets have been shown to be ubiquitous in the Milky Way, and most have been discovered with two primary methods: the transit method and the radial velocity method. These techniques are complementary to each other; the transit method identifies the radius of a given exoplanet while the radial velocity method can be used to derive its mass. Space missions to carry out the transit method such as \textit{Kepler} \citep{kepler2010} and the recently launched \textit{Transiting Exoplanet Survey Satellite} \citep[TESS,][]{ricker2014} have dominated new exoplanet discoveries, with thousands revealed in recent years and many more to come with each release of TESS data. Even with the success of these space missions, the radial velocity method is paramount to ground-based followup efforts to confirm the exoplanets and their masses, and many new ground-based radial velocity instruments are planned or have been recently commissioned \citep{wright2017}. Furthermore, transiting exoplanets with mass measurements are ideal for atmospheric characterization studies, either with ground-based high-resolution spectroscopy or space-based low-resolution spectroscopy with missions such as the upcoming \textit{James Webb Space Telescope}.

With instrument precision approaching 10 cm s$^{-1}$, the fidelity of spectroscopic data makes it possible to resolve astrophysical velocity sources in stellar atmospheres and micro-telluric contamination from Earth's atmosphere. These two error sources now limit the total measurement precision attainable with the radial velocity method. New analysis techniques are being developed to understand and mitigate the impact of stellar activity effects \citep[e.g.,][]{haywood2014,davis2017,dumusque2018}, and telluric contamination \citep{seifahrt2010,wise2018,bedell2019,leet2019}. With improved instrumentation, these techniques will be paramount to reduce the total measurement uncertainty to the goal of 10 cm s$^{-1}$, which will enable the discovery of Earth-like exoplanets \citep{fischer2016}. 

One crucial step in the radial velocity method is the barycentric correction, which is to account for the velocity of the Earth relative to the barycenter of the solar system. This paper follows up on \cite{blackman2017}, which predicted the impact of variable chromatic atmospheric attenuation on barycentric corrections, and suggested the use of a low-resolution spectrograph as an exposure meter for radial velocity instruments. Such a system has now been built and commissioned with the EXtreme PREcision Spectrograph \citep[EXPRES;][]{jurgenson2016}. Here, we present the empirical magnitude of the radial velocity error if wavelength-dependence in the barycentric correction is ignored. In Section \ref{sec:2}, we describe the implementation of the chromatic exposure meter of EXPRES. In Section \ref{sec:3}, observational results are presented for a single case as well as the entire ensemble of EXPRES observations thus far. In Section \ref{sec:4}, we discuss how the impact of chromatic atmospheric effects depends on the method used to solve for the stellar radial velocity, as well as other parameters such as the instrument and observation strategy. In Section \ref{sec:5}, we summarize our results.

\section{Hardware Setup and Data Reduction}
\label{sec:2}
\subsection{Overview of EXPRES}
EXPRES is a new radial velocity instrument for exoplanet discovery and characterization, recently commissioned at the 4.3 meter Discovery Channel Telescope at Lowell Observatory. Designed to discover rocky exoplanets in the solar neighborhood, EXPRES is an environmentally stabilized, cross-dispersed echelle spectrograph operating at visible wavelengths with a resolving power reaching 150,000. The design driver for EXPRES is to have the resolution and instrumental precision necessary to isolate and remove the effects of stellar activity from observed spectra. This correction would isolate the Doppler signatures of orbiting exoplanets. To achieve this goal, many novel features have been implemented on EXPRES, based on our analysis of weaknesses in previous instruments. A Menlo Systems laser frequency comb (LFC) is used as the primary wavelength calibration source \citep[e.g.,][]{wilken2012,molaro2013,probst2014} with a thorium-argon lamp used for initial, coarse wavelength solutions. Flat-field calibration is performed with a dedicated fiber that is larger than the science fiber, providing higher SNR at the edges of the echelle orders. The flat-field light source is a custom, LED-based source that is inversely tuned to the instrument response. Modal noise in the multimode fibers is mitigated with a chaotic fiber agitator \citep{petersburg2018}. Finally, the chromatic exposure meter enables wavelength-dependent barycentric corrections. 

\subsection{Exposure meter design}
The EXPRES exposure meter is composed of a commercially available Andor iXon 897 electron-multiplying charge-coupled device (EMCCD) and an Andor Shamrock 193i Czerny-Turner spectrograph. This spectrograph has a focal length of 193 mm, a ruled grating with 150 $\mathrm{lines}/\mathrm{mm}$ and 500 nm blaze, and a resolving power peaking at  $R\approx 100$. The resolution has been empirically measured using lasers of various wavelengths and two bright argon lines in the red from the thorium-argon lamp. No slit is used to retain as much light as possible, at the expense of spectral resolution. The bandpass of the spectrograph can be adjusted, and is matched to the wavelength range of the LFC, 450 nm to 710 nm.  This instrument is fed by a 200 $\mu$m circular optical fiber, which receives light from a 2\% beam splitter within the EXPRES vacuum chamber just before light is injected into the main spectrograph optics. The re-imaging of science light into this fiber is relatively efficient, as the rectangular science fiber core is smaller at 180 $\mu$m $\times$ 33 $\mu$m. The throughput of the exposure meter spectrograph has a peak value of roughly 45\% at 600 nm. Given the coupling efficiency between the science and exposure meter fibers, and the relative throughput of EXPRES and its exposure meter, we estimate that flux to the exposure meter EMCCD is about 2\% of that on the EXPRES CCD. This flux is typically sufficient for one-second integrations with the exposure meter EMCCD for stars up to $V = 8$, depending on seeing. Fainter stars may be observed with increased integration lengths of the EMCCD.

The shutter of EXPRES is located in a pupil slicer and double scrambler module that is spliced into the science fiber, before the vacuum chamber. Therefore, a single shutter controls light flow to both EXPRES and its exposure meter. The shutter is controlled by a National Instruments CompactRIO controller that is constantly syncing its clock to absolute sources, as the reported shutter open and close times are required to be accurate to 0.25 seconds in order to calculate barycentric corrections with errors less than 1 cm s$^{-1}$ \citep{we2014}.

\subsection{Exposure meter data reduction}
For each one-second exposure meter integration, a full $512\times512$ pixel array is read out from the detector. While stellar light is only recorded on the central 30 pixels in the cross-dispersion direction, the surrounding regions are used for a dark and bias subtraction of the chip that is interpolated over the spectral region. Cosmic ray rejection is performed with the L.A.Cosmic algorithm \citep{vandokkum2001}. Wavelength calibration was originally performed by injecting lasers of several different wavelengths into the spectrograph, and can be periodically checked and adjusted via thorium-argon spectra that are regularly taken as part of the EXPRES nightly calibration procedures. The $512\times30$ pixel spectra are boxcar extracted, and saved along with the geometric midpoint of each integration of the EMCCD. The geometric midpoint time of each exposure meter integration is found by extrapolating from the EXPRES shutter open and close times. The exposure meter spectra are then binned into discrete wavelength channels. The effective wavelength of each channel is found by taking a photon-weighted average of the wavelengths present in that channel. A photon-weighted barycentric correction is computed for each channel, and these corrections are fit with a low-order polynomial to interpolate over all wavelengths of the EXPRES spectrum. The number of channels and fitting function can be changed arbitrarily depending on the nature of the data, but eight channels and a third-order polynomial is typically appropriate for the EXPRES data. This procedure is described in more detail in \cite{blackman2017}.

\subsection{Application of the barycentric correction}
The barycentric correction algorithm used in this analysis is that of \cite{kanodia2018}. This algorithm is ideal for exoplanet searches with the radial velocity method, given that its maximum error is under 1 cm s$^{-1}$ (assuming accurate inputs), and has been tested for accuracy against previous standards \textit{zbarycorr} \citep{we2014} and \textit{TEMPO2} \citep{hobbs2006}. The barycentric correction is applied with the formulation of \cite{we2014}, modified to include a wavelength-dependence, via
\begin{equation}
z_{true} = [z_B(\lambda)+1][z_{obs}(\lambda)+1]-1,
\end{equation}
where $z_{true}$ is the Keplerian, barycentric corrected radial velocity of the star, $z_B(\lambda)$ is the barycentric correction velocity which can be different for different wavelengths, and $z_{obs}(\lambda)$ is the observed stellar radial velocity as a function of wavelength. The quantity $z_{obs}(\lambda)$ may have wavelength dependence from both astrophysical and atmospheric effects. We solve for the approximate Keplerian radial velocity of the star with the cross-correlation function (CCF) method. This quantity does not have a wavelength dependence, however, the computed radial velocity of the star may have a small wavelength dependence even after barycentric correction, as different wavelengths inherently probe different surfaces of the star. We do not attempt to measure such a dependence. The CCF is computed for each spectral order separately, and these are co-added. A Gaussian-like model is then fit to the co-added CCF, where the radial velocity of the star is the mean of the model. Therefore, we do not actually obtain $z_{obs}(\lambda)$ with this method, but instead solve for the best-fit velocity for the input spectrum all at once. The barycentric correction is performed as a wavelength shift of each pixel before this CCF is computed, and the resulting radial velocity is in the frame of the solar system barycenter.

\section{Observational Results}
\label{sec:3}
\subsection{Example of atmospheric effects in a single observation}
\begin{figure*}[]
\begin{center}
\includegraphics{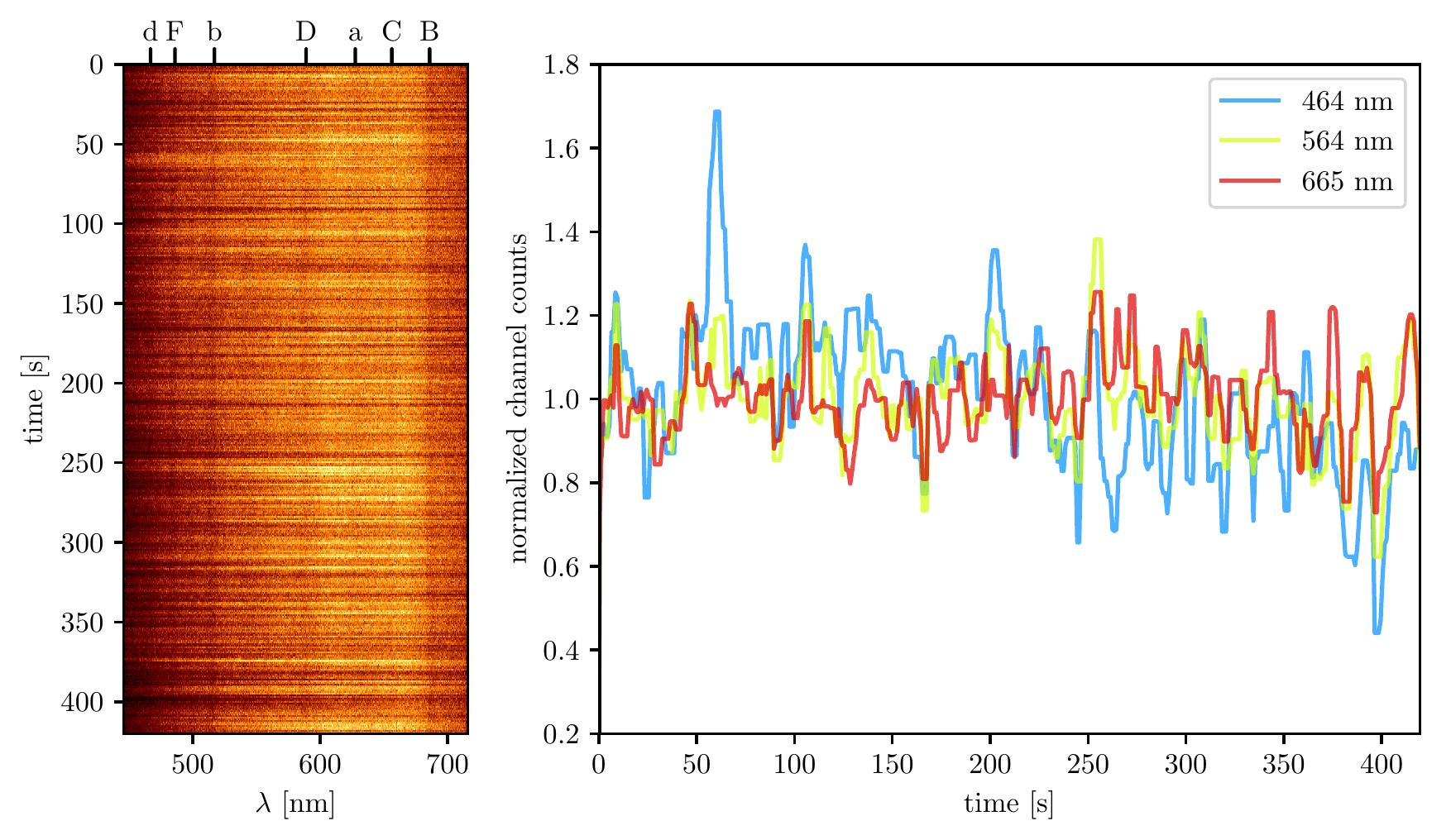} 
\end{center}
\caption{\textit{Left:} An example time series of 51 Pegasi spectra from the EXPRES exposure meter. Significant chromatic changes in flux are observed over the course of this observation. The locations of visible absorption lines are labeled at the top. \textit{Right:} Normalized count rate of blue, green, and red channels for the same observation. Blue photon flux was stronger at the beginning of the exposure and decreased toward the end. A median filter with kernel size of 5 was applied to this data so that the effect can be more clearly seen.}
\label{fig:expm_data}
\end{figure*}

In the left panel of Figure \ref{fig:expm_data}, we show the full exposure meter data set for one 420 second observation of 51 Pegasi. Each exposure meter integration is one second in duration, which results in 420 individual exposure meter spectra for this observation. Each row in the image corresponds to one extracted exposure meter spectrum that has been bias and dark subtracted with cosmic rays removed. The spectra are 512 pixels wide and cover wavelengths from 450 nm to 710 nm. This matches the spectral range of the laser frequency comb used for EXPRES wavelength calibration and the wavelength range used for the radial velocity analysis. There are several important features to note in this figure. First, the apparent brightness of the star varies from integration to integration, which could be caused by a number of effects, such as seeing, cloud cover, or guiding errors. The second feature is that these changes in brightness are not equal at all wavelengths. For example, at 60 seconds, there is an apparent excess of blue photons without a corresponding increase in red photons. The vertical dark regions are absorption lines labeled in Fraunhofer notation from either the stellar spectrum (d, F, b, D,  and C) or strong telluric lines (a and B) from Earth's atmosphere. In the right panel of Figure \ref{fig:expm_data}, we show the normalized count rate for three of the eight channels used in this exposure. The apparent spike in blue photon flux can be seen, as well as an overall decrease in counts in blue wavelengths throughout the exposure. A median filter with kernel size 5 was applied to these data so that these effects were more apparent, but such a filter is not used otherwise in the analysis.

\begin{figure}[]
\begin{center}
\includegraphics{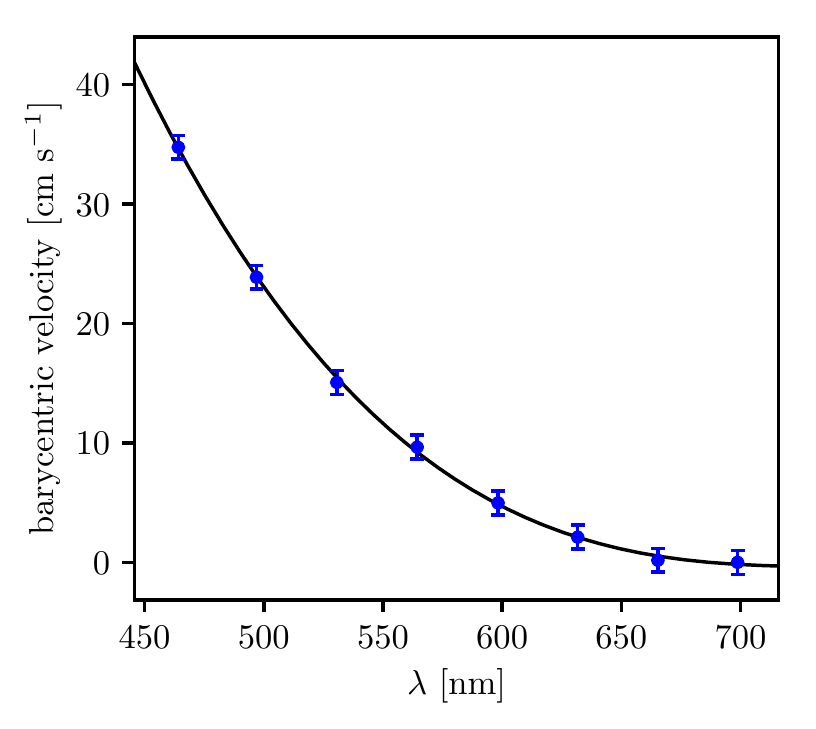} 
\end{center}
\caption{The barycentric correction velocities fit with a third-order polynomial for the observation shown in Figure \ref{fig:expm_data}. A significant chromatic effect is observed as the corrections at the blue and red ends of the spectrum differ by 40 cm s$^{-1}$.}
\label{fig:bc_lambda}
\end{figure}

For this observation, the resulting barycentric correction velocities as a function of wavelength are plotted in Figure \ref{fig:bc_lambda}.  A third-order polynomial is fit to the different channel barycentric corrections in order to interpolate a correction for all wavelengths. The barycentric correction exhibits a wavelength-dependence, as it differs by 40 cm s$^{-1}$ from one end of the spectrum to the other. This equates to a 12.2 cm s$^{-1}$ radial velocity error, assuming the CCF method of solving for radial velocity discussed previously. The details regarding how this error is calculated are discussed in the next section. The uncertainty in each barycentric correction velocity is estimated from several sources of error, including the accuracy of the reported shutter open and close times, the accuracy of the algorithm used to calculate the barycentric correction, the accuracy of the astrometric solution of the target star, and the limited signal-to-noise ratio (SNR) of the exposure meter spectra. The formal errors from the first two error sources are close to 1 mm s$^{-1}$. The astrometric solution for each target star has been determined with \textit{Gaia} to an accuracy of tens of $\mu$as for the position, parallax, and proper motion, rendering this error source negligible \citep{we2014,gaia2016}. For the relatively bright stars in this radial velocity survey, the SNR of the exposure meter spectra is high ($>100$). Taking these factors into account, we conservatively estimate the uncertainty in the barycentric correction at 1 cm s$^{-1}$ for each channel, which matches the typical allotment for the barycentric correction error in the error budgets of high-precision radial velocity instruments \citep[e.g.,][]{podgorski2014,halverson2016,jurgenson2016}.

While this observation and other observations of 51 Pegasi from the same night exhibit a strong chromatic dependence in the barycentric correction, on other nights, observations of the same star, at the same air mass and observation length, exhibit much smaller effects. This night-to-night variability is expected if atmospheric conditions are the primary cause of the chromatic flux changes.

\subsection{Results from all EXPRES observations}
\begin{figure}[ht]
\begin{center}
\includegraphics{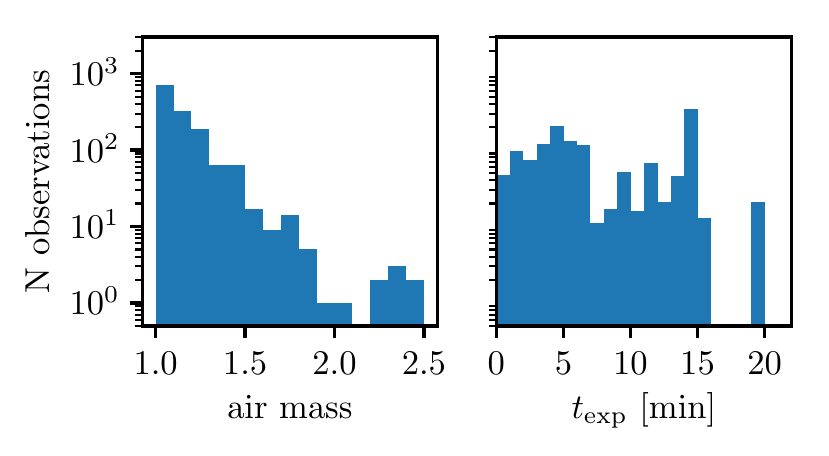} 
\end{center}
\caption{Histograms of the air mass values and observations lengths for the EXPRES observations in this sample. The air mass bins are 0.1 in width and the exposure length ($t_{\mathrm{exp}}$) bins are 60 seconds in width.}
\label{fig:am_explengths}
\end{figure}

With hundreds of completed EXPRES observations under a variety of atmospheric conditions over a one-year period, we now determine the typical significance of chromatic dependance in the barycentric correction from a large sample. The number of unique stars observed is approximately 100. The distributions of air mass and observation length for all EXPRES observations in this period are shown in Figure \ref{fig:am_explengths}. 

\noindent

For the purpose of determining the typical radial velocity error incurred by chromatic atmospheric effects, we only consider exposure lengths greater than 250 seconds in the remaining analysis. These are typical exposure lengths for radial velocity observations of exoplanet host stars with high-resolution spectrographs. Some shorter observations have been performed with EXPRES, and those targets are typically bright B-type stars, which are used for telluric contamination analysis. In the subset of 1064 observations longer than 250 seconds, 96.1\% were performed at an average air mass below 1.5, 34.9\% of exposures were 15 minutes or greater in duration, and the longest exposures were 20 minutes. Most stars in this analysis are bright ($V<8$), and the expected radial velocity error contribution from photon noise is low ($< 50$ cm s$^{-1}$).

\begin{figure*}[]
\begin{center}
\includegraphics{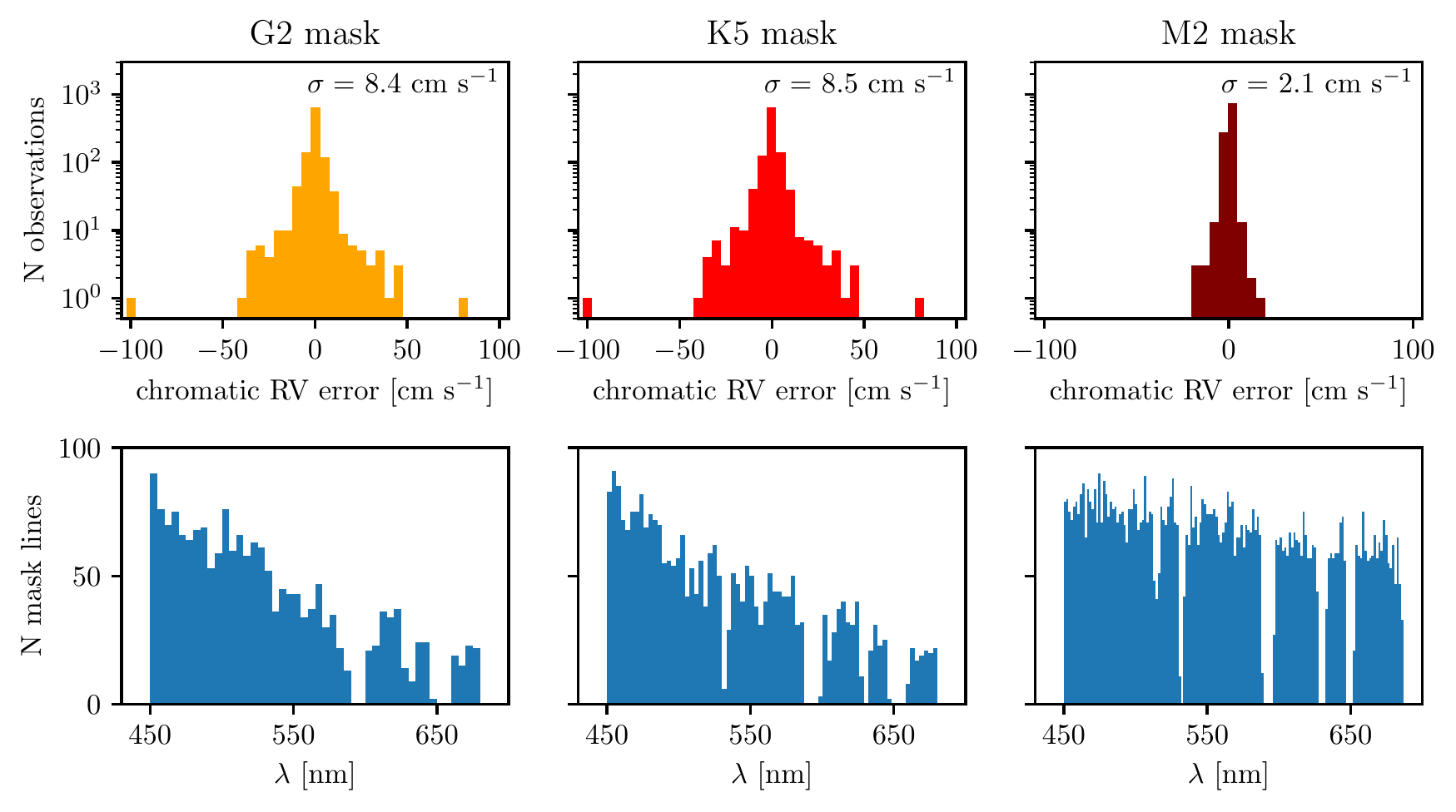}
\end{center}
\caption{\textit{Top row}: The distributions of chromatic radial velocity (RV) errors for three different spectral masks used in the CCF method of calculating stellar radial velocity. \textit{Bottom row}: The distribution of stellar absorption lines from the corresponding mask.}
\label{fig:3hists}
\end{figure*}

With the chromatic exposure meter data for each EXPRES observation, we calculate what the incurred radial velocity error would be if the chromatic component of the barycentric correction is ignored, assuming three different spectral masks for the CCF method of solving for stellar radial velocity. These masks have been developed for G2, K5, and M2 spectral types, and are inherited from the CERES package for reducing high-resolution spectra \citep{brahm2017}. In calculating the chromatic radial velocity error, we have included the line weights from each mask. These are the errors that would be incurred if a single channel exposure meter was used for the EXPRES observations, as is the case for previous radial velocity instruments such as HARPS, HIRES, and CHIRON  \citep{mayor2003,kibrick2006,tokovinin2013}. Figure \ref{fig:3hists} shows histograms of the distributions of incurred errors for each of the three mask types. Below each error distribution, a histogram shows where the mask lines occur in the spectrum for the respective mask. A summary of the errors incurred for each mask type is shown in Table \ref{tab:errors}. The largest chromatic radial velocity error is 1 m s$^{-1}$, and a majority of the chromatic radial velocity errors for G2 and K5 masks are greater than 1 cm s$^{-1}$. The dependence on mask type is caused by the distribution of lines present in the mask. G2 and K5 masks have a much higher density of lines in blue wavelengths compared to red wavelengths. This increases the significance of the chromatic dependence in the barycentric correction, as preferentially using lines from one side of the spectrum prevents the chromatic effect from averaging out. The distribution of absorption lines in the G2 and K5 masks are similar, and so the incurred radial velocity errors are also similar. The distribution of absorption lines in the M2 mask is much more uniform, and so the incurred error is smaller.

\begin{table}[h]
\caption {Summary of radial velocity errors incurred with each stellar mask type used in the generation of the CCF. The first column shows the standard deviation of errors, and the second and third columns show the percentage of errors greater than 1 cm s$^{-1}$ and 10 cm s$^{-1}$, respectively.} \label{tab:errors} 
\centering
\begin{tabular}{l|lll}
\hline
\hline
   & $\sigma$  & $\epsilon_\lambda>1$ cm s$^{-1}$ & $\epsilon_\lambda>10$ cm s$^{-1}$ \\ \hline
G2 & $8.4$ cm s$^{-1}$ & $62.4\%$                       & $9.9\%$                        \\
K5 & $8.5$ cm s$^{-1}$ & $62.7\%$                       & $9.7\%$                        \\
M2 & $2.1$ cm s$^{-1}$ & $22.4\%$                       & $0.9\%$                        
\end{tabular}
\end{table}
\noindent

To be exceedingly accurate, it is possible to exclude regions of telluric absorption lines from the exposure meter analysis. Such wavelengths are not technically used in computing the radial velocity, and so atmospheric variability there is not relevant for the barycentric correction. We explored this possibility by excluding the two gaps in the G2 mask CCF lines from the exposure meter spectra, and then we refit all of the chromatic barycentric corrections. The differences in the resulting barycentric correction velocities are typically negligible, at a level $< 1$ mm s$^{-1}$, and in the worst cases, around 0.25 cm s$^{-1}$. This low impact can be explained by two factors. First, these regions are relatively narrow, and probably do not have a large impact on the varying flux in the barycentric correction channels that contain them. Furthermore, the chromatic variability that we see in this study is fairly smooth across optical wavelengths, and any flux changes in these telluric regions is similar in surrounding wavelengths.

\begin{figure*}[]
\begin{center}
\includegraphics{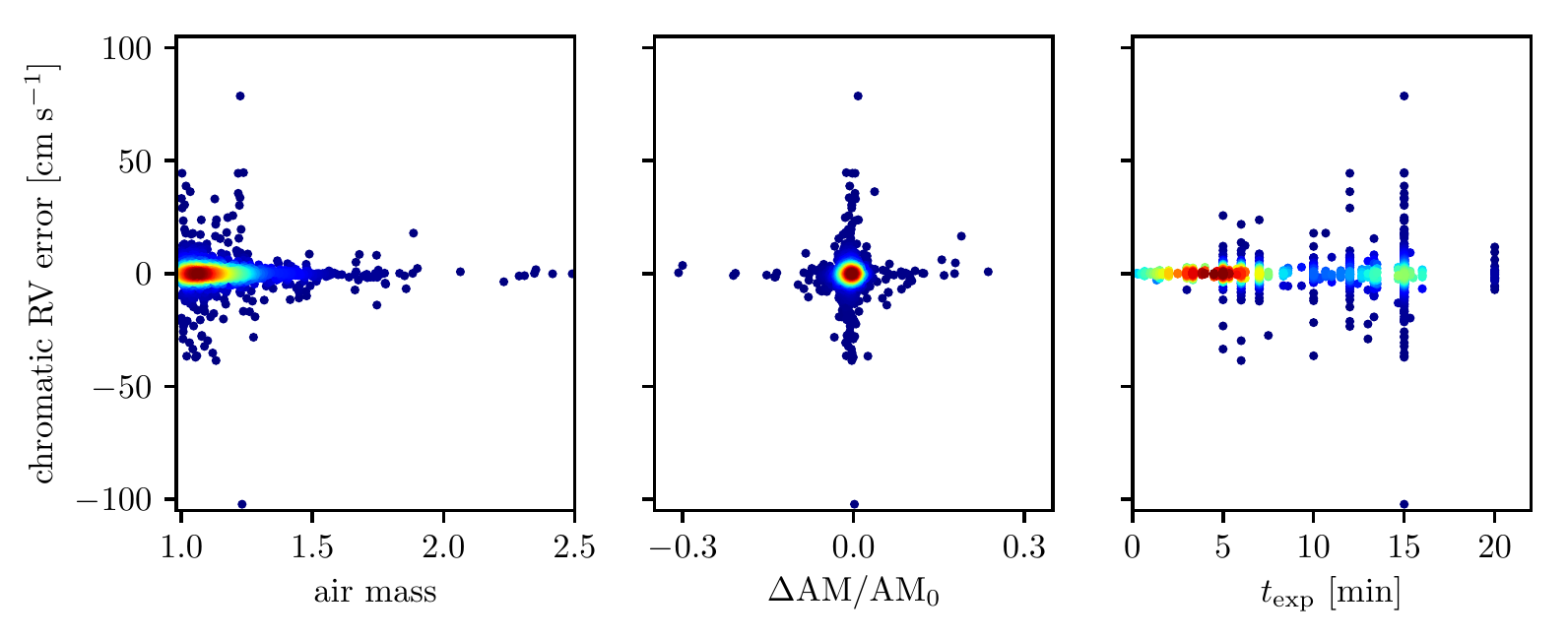} 
\end{center}
\caption{Chromatic radial velocity error plotted against air mass, fractional change in air mass, and exposure length from results assuming the G2 CCF mask. The errors do not correlate with air mass or fractional change in air mass, and longer observations are more susceptible to large chromatic errors. The points are colored by density in the parameter space, scored by a Gaussian kernel density estimation, with red as higher density, and blue as lower density.}
\label{fig:error_vs_am_explengths}
\end{figure*}

The observation parameters, such as air mass, fractional change in air mass, and exposure length, are of interest in cases of large chromatic dependence in the barycentric corrections. The chromatic radial velocity errors are plotted against these quantities in Figure \ref{fig:error_vs_am_explengths}. In this Figure, we include observations shorter than 250 seconds in duration as well, in the interest of searching for correlations with exposure length.

\cite{blackman2017} noted that air mass changes more rapidly at higher air mass values when observing in the east or west. Such changes in air mass induce a ubiquitous chromatic dependence in the barycentric corrections due to changes in the atmospheric transmission spectrum. However, at high air masses for a star near to the meridian, the air mass does not change rapidly during an observation. Therefore, air mass alone may not be a strong predictor of the strength of chromatic errors in the barycentric correction, because azimuth also determines what the change in air mass will be for a given observation. In Figure \ref{fig:error_vs_am_explengths}, we see a roughly uniform distribution of errors for air masses greater than 1.5. At lower air masses, the chromatic error appears to be larger, and also roughly uniform. The vast majority of observations were performed at low air mass, it may therefore be an observational bias that the largest errors are observed in these cases. In the center panel of Figure \ref{fig:error_vs_am_explengths}, we see a slight anti-correlation with fractional change in air mass; the largest errors occurred during small changes in air mass while small errors were incurred at large changes in air mass. Again, this is most likely due to observational bias, as we have not evenly sampled the range of fractional changes in air mass.

The right panel in Figure \ref{fig:error_vs_am_explengths} shows a rough dependence of chromatic error on exposure length, but there are exceptions. This is expected, a longer observation will give a star more time to traverse the sky and more time for the composition of the atmosphere to change along the line of sight. However, long exposure lengths are not a guarantee that large chromatic errors will be incurred. If the atmosphere is stable and air mass is low, the chromatic effects should be small. This could have been the case for our 15 and 20 minute exposures which exhibited small chromatic effects. Shorter exposures did tend to incur smaller errors, as no exposure shorter than 5 minutes incurred an error over 5 cm s$^{-1}$ in our results. 

\begin{figure*}[]
\begin{center}
\includegraphics{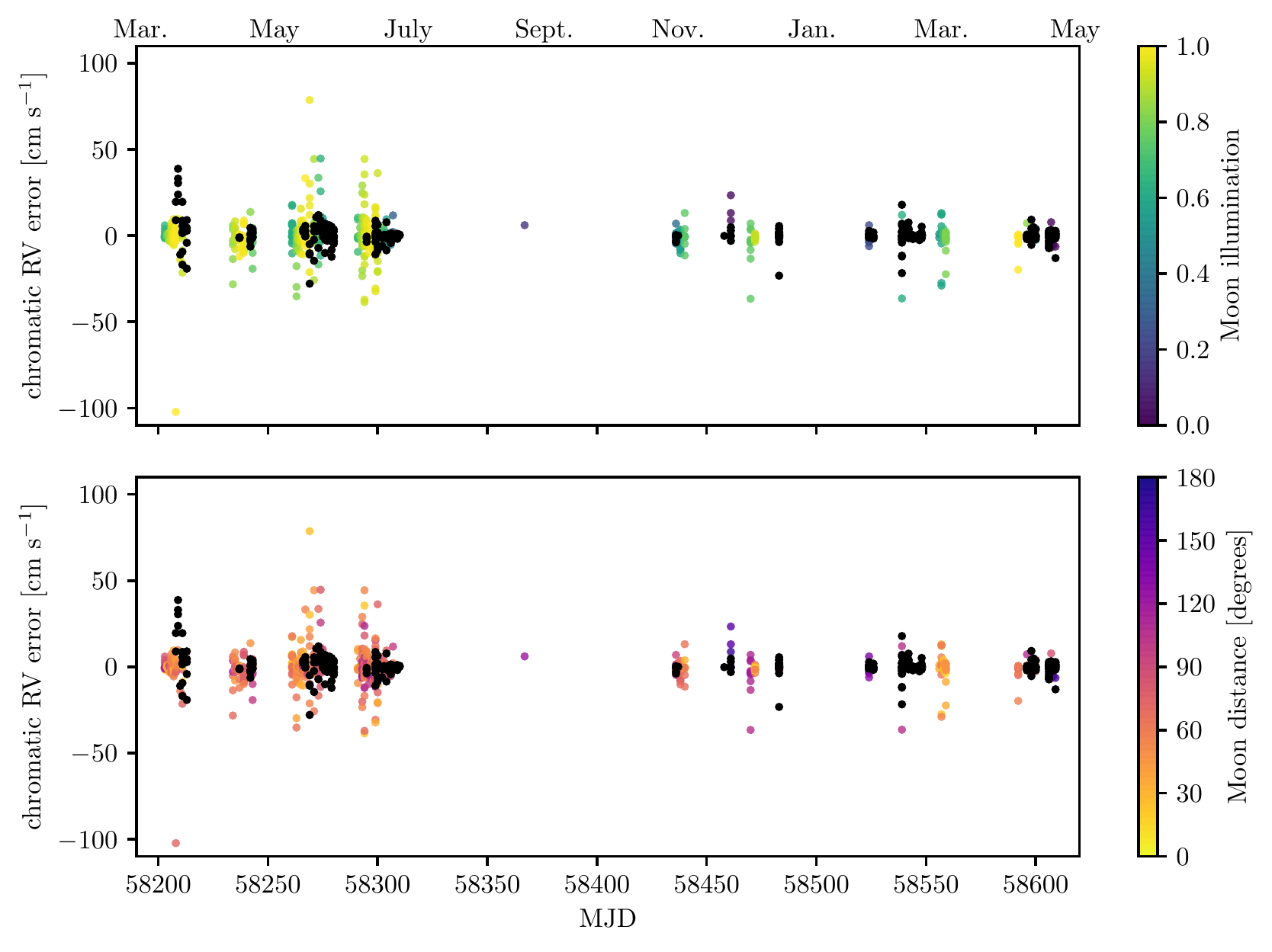}
\end{center}
\caption{\textit{Top:} the chromatic radial velocity errors assuming the G2 CCF mask, grouped by date observed, and colored by the phase of the Moon, with black corresponding to the Moon being below the horizon. Larger errors tend to be observed on the same nights, and occur throughout the year. No significant correlation with season or phase of the Moon is observed. \textit{Bottom:} the same data as above, colored by angular distance to the Moon.}
\label{fig:error_vs_night}
\end{figure*}

Without strong correlations with known observing parameters, we now examine whether the chromatic errors tend to be similar on a given night. In Figure \ref{fig:error_vs_night}, we plot the chromatic radial velocity errors as a function of the night that they were observed on. In the top panel, points are colored by fraction of the Moon that is illuminated, except in cases when the moon is below the horizon, the points are colored black. In the bottom panel, points are colored by angular distance to the Moon. With just a few exceptions, large chromatic errors tend to occur on nights with other large chromatic errors. No correlation with season, phase of the Moon, or angular distance to the Moon is observed. Some of the largest errors did occur when the Moon was nearly full in the sky, and one concern is that moonlight may be reflected off of clouds in a variable way, contaminating these observations. However, large errors also occurred on nights with no visible moon, and there were other nights with an illuminated moon that exhibited small chromatic errors. This result suggests that nightly observing conditions are the primary determinant to whether strong chromatic effects will be present. For example, cloud cover, wavelength-dependent seeing, and atmospheric composition, such as the presence of water vapor and aerosols, may change on short timescales. Furthermore, the spectral energy distributions of stars do not vary significantly on such short timescales, and no known instrument effect would cause chromatic changes in throughput, although the possibility of additional error due to a lack of atmospheric dispersion correction is discussed in section \ref{sec:4.2}.
 
\section{Discussion}
\label{sec:4}
The exact radial velocity error induced by chromatic atmospheric effects on any given observation depends on many factors, which can be grouped into three categories:

\begin{itemize}
\item the method used to solve for the stellar radial velocity
\item specific details of the instrumentation used
\item the parameters of the observation, including the site characteristics
\end{itemize}

\subsection{Alternate methods of solving for radial velocity}
The results presented thus far have assumed the CCF method of solving for stellar radial velocity. In this method, a weighted binary mask containing the wavelengths of absorption features deemed to be suitable for radial velocity measurements is stepped across a wavelength-calibrated stellar spectrum at a certain velocity interval \citep[e.g.,][]{baranne1979,pepe2002}. At each velocity location, the observed spectrum is summed in the binary mask. The result is a CCF that traces the average spectral line profile, which can be fit with a Gaussian-like model, where the mean of the model is the radial velocity of the star. The nature of the chromatic errors on barycentric corrections is impacted in this method by the number and distribution of mask lines, as seen in Figure \ref{fig:3hists}. A line-by-line analysis \citep[e.g.,][]{dumusque2018} will be similarly susceptible to these chromatic effects for G-type and K-type stars, as the distribution of suitable absorption features for radial velocity measurement is heavily weighted towards the blue end of the spectrum of G-type and K-type stars.

\begin{figure}[h]
\begin{center}
\includegraphics{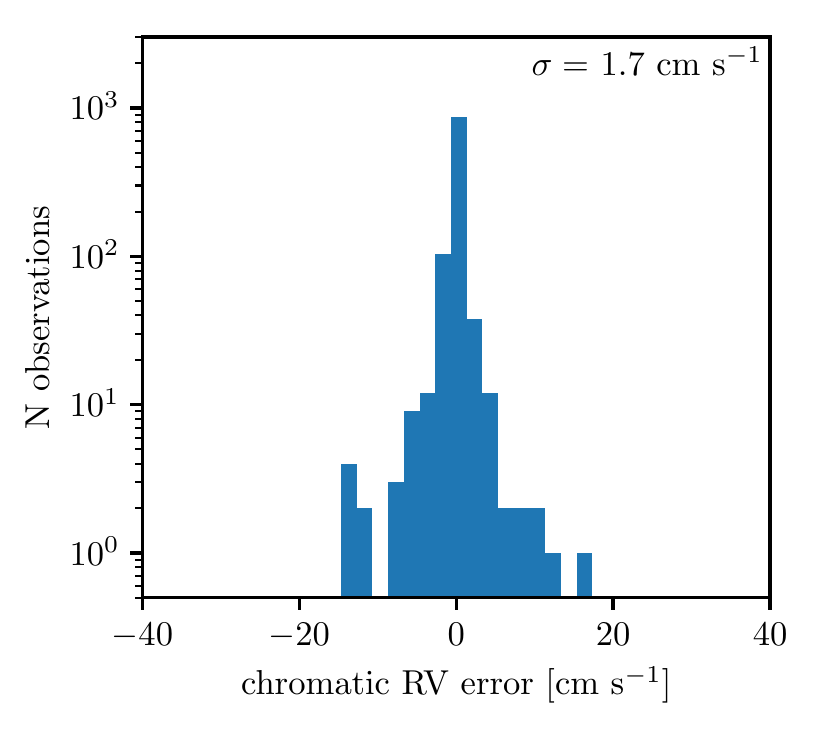} 
\end{center}
\caption{The distribution of radial velocity errors from chromatic atmospheric effects assuming the least squares template matching method of computing radial velocity. The impact is less significant in this scenario because the information used for computing the radial velocity is more uniformly distributed across the spectrum.}
\label{fig:uniform_hist}
\end{figure}

Another approach to solving for radial velocity is to combine multiple observations of a given star to create a high-SNR template spectrum, and match that to each observation by adding a velocity shift \citep[e.g.,][]{anglada2012,zechmeister2018}. The radial velocity of the star is then taken to be the best-fit velocity offset with a least-squares method. With this method, all pixels of the spectrum are weighted by SNR, and regions with telluric lines are masked out. When this occurs, the impact of the chromatic atmospheric effects on barycentric corrections is minimized, as the barycentric corrections across the spectra are typically symmetric about the central wavelength. If equal an amount of information is being used from each side of the central wavelength, the differences in the barycentric correction tend to average out. This is analogous to case of the M2 CCF mask described in Section \ref{sec:3}. In Figure \ref{fig:uniform_hist}, we show the distribution of chromatic radial velocity errors that would be incurred with this method. The standard deviation of radial velocity errors is 1.7 cm s$^{-1}$, more than a factor of four smaller than with the CCF method for G-type and K-type stars, and only slightly smaller than the errors for M-type stars with the CCF method. It may then be expected that the template matching method of \cite{anglada2012} would produce better results for G-type and K-type stars due to being less susceptible to chromatic atmospheric effects, however, significant improvement in radial velocity precision was only found for M-type stars. This was likely due to extraneous factors, notably the additional continuum noise of G-type and K-type stars. This method is not currently being used for EXPRES observations, as there can be benefit in selecting specific absorption lines. For example, lines can be selected based on depth or susceptibility to stellar activity, and assigned individual weights.

\subsection{Considerations from instrumentation}
\label{sec:4.2}
The design of the spectrograph used to measure stellar radial velocity will affect the impact of the chromatic atmospheric effects. The primary quantity of interest is the wavelength range of the region used for radial velocity analysis. The broader the wavelength range, the larger the effect will be from one end of the spectrum to the other. For example, the iodine region of the CHIRON spectrograph \citep{tokovinin2013} spanned 500 nm to 600 nm. A green light filter was placed in front of the single-channel photomultiplier tube exposure meter, which limited the exposure meter flux to this 100 nm region. While there was no wavelength information in the exposure meter, the magnitude of the chromatic effects would have been limited, owing to the relatively narrow range of wavelengths used in the Doppler analysis. Spectrographs that cover the entire optical window use a wavelength region that is about three times larger, thus are more susceptible to chromatic atmospheric effects. This may be especially relevant for instruments extending further into blue wavelengths, as we have observed that $z_B(\lambda)$ is typically steepest in blue wavelengths. The wavelength coverage for radial velocity measurements with EXPRES starts at 450 nm.

A second important aspect of the instrument is the guiding method and use of atmospheric dispersion correction (ADC). Without ADC, the image of the star will be elongated with a chromatic dependence. Some observations were excluded from this analysis because of documented instrument problems during observations; large chromatic effects were observed when there were problems with the ADC and the fast tip-tilt (FTT) guiding system of EXPRES. For example, on one night, the FTT system was left off, and the guiding of starlight onto the fiber was performed manually. When the star position was adjusted during an exposure, this caused a nearly instantaneous chromatic change in the recorded spectrum. At this time, the ADC was not yet performing optimally, and guiding on a different part of the star had a large chromatic impact. Therefore, proper atmospheric dispersion correction and guiding could be paramount to limiting chromatic atmospheric effects during observations.

We also note that the exposure meter integration frequency will impact the degree to which chromatic flux changes can be measured. This variability occurs on timescales of seconds in the EXPRES exposure meter data. If longer integrations are used to increase SNR in the exposure meter, there is considerable risk in not detecting the full extent of the flux changes. We have binned the EXPRES exposure meter data in time to simulate this effect; integration times longer than 40 seconds lead to radial velocity errors greater than $1$ cm s$^{-1}$. Considering this, we note that the 2\% beamsplitter for the exposure meter has been appropriate for one-second exposure meter integrations on the relatively bright targets observed by EXPRES. On fainter targets, this integration length would need to be increased or else the beamsplitter would need to pick off a larger ratio of light for the exposure meter. This design decision also depends on the telescope size and instrument throughput up to the beamsplitter, future instrument designs will need to consider these factors as well.

\subsection{Observing parameters}
The magnitude of the chromatic effects depends on the observation parameters, and these include stellar spectral type, length of the observation, and transient changes in atmospheric composition during the observation. The last of these effects is both the most significant and the most unpredictable, with a dependence on site location, which dictates the need for an exposure meter during every observation. Explanations of some potential effects and their significance were presented in \cite{blackman2017}. For example, the Discovery Channel Telescope is located in a forest in which there are often controlled burns and wildfires, these events produce smoke with particle sizes similar to visible wavelengths. We have also measured night-to-night changes in the amount of precipitable water vapor in the atmosphere. Variability in these quantities along the line of sight during observations could manifest as wavelength-dependent flux changes detectable by the exposure meter. We note that if the results presented here are to be extrapolated to estimate chromatic errors with other instruments, the instrumentation, observatory site, and the lengths of the observations should be carefully considered. Longer observations will risk larger chromatic errors, and shorter observations will limit them.

\section{Conclusion}
\label{sec:5}
In order to reach a measurement precision goal of 10 cm s$^{-1}$ with the radial velocity method for exoplanet discovery and characterization, every source of instrumental error must be understood and mitigated. A low-resolution exposure meter spectrograph commissioned with the EXPRES instrument has been used to measure typical wavelength-dependent changes in atmospheric transmission along the line of sight during observations. The photon-weighted barycentric correction must be performed as a function of wavelength to account for chromatic variability in stellar flux during observations. If the chromatic dependence in the barycentric correction is ignored, radial velocity errors exceeding 10 cm s$^{-1}$ can be incurred. This error depends on the atmospheric conditions, the method used in solving for the stellar radial velocity, specific details of the instrumentation, and the observing parameters. For bright stars, this error is of comparable size to that expected from photon noise and instrumental effects. Therefore, chromatic atmospheric effects are important to mitigate for instruments attempting to reach the highest possible radial velocity measurement precision.

\software{Barycorrpy \citep{kanodia2018}, CERES \citep{brahm2017}, Astropy \citep{astropy2013,astropy2018}, Matplotlib \citep{matplotlib2007}, SciPy \citep{scipy2001}, NumPy \citep{numpy2011}}

\acknowledgments
We thank the anonymous referee for providing valuable suggestions that improved this manuscript.
Support for this work was provided by the National Science Foundation under grant NSF MRI 1429365.
We also thank Ren{\'e} Tronsgaard and Didier Queloz for many fruitful discussions about barycentric corrections.
 
\bibliographystyle{aasjournal.bst}

\end{document}